\documentclass[aip, apl, reprint, longbibliography]{revtex4-1}
\usepackage{graphicx, hyperref, amsmath, amsfonts, amssymb, amsthm, mathtools}

\hypersetup{
colorlinks=true,
linkcolor=blue,
filecolor=magenta, 
urlcolor=cyan,
citecolor=cyan
}

\begin{document}

\title{Biorealistic response in a technology-compatible graphene synaptic transistor}

\author{Anastasia Chouprik}
\email[]{chouprik.aa@mipt.ru}
\affiliation{Moscow Institute of Physics and Technology, Dolgoprudny, Russia}

\author{Elizaveta Guberna}
\affiliation{Moscow Institute of Physics and Technology, Dolgoprudny, Russia}

\author{Islam Mutaev}
\affiliation{Moscow Institute of Physics and Technology, Dolgoprudny, Russia}

\author{Ilya Margolin}
\affiliation{Moscow Institute of Physics and Technology, Dolgoprudny, Russia}

\author{Evgeny Guberna}
\affiliation{Moscow Institute of Physics and Technology, Dolgoprudny, Russia}

\author{Maxim Rybin}
\affiliation{Prokhorov General Physics Institute of the Russian Academy of Sciences, Moscow, Russia}

\begin{abstract}
Artificial synapse is a key element of future brain-inspired neuromorphic computing systems implemented in hardware. This work presents a graphene synaptic transistor based on all-technology-compatible materials that exhibits highly tunable biorealistic behavior. It is shown that the device geometry and interface properties can be designed to maximize the memory window and minimize power consumption. The device exhibits a virtually continuous range of multiple conductance levels, similar to synaptic weighting, which is achieved by gradual injection/emission of electrons into the floating gate and interface traps under the influence of an external electric field. Similar to the biological synapse, the transistor has short-term intrinsic dynamics that affect the long-term state. The temporal injection/emission dynamics of an electronic synapse closely resemble those of its biological counterpart and is exploited to emulate biorealistic behavior using a number of synaptic functions, including paired-pulse facilitation/depression, spike-rate-dependent plasticity, and others. Such a synaptic transistor can serve as a building block in hardware artificial networks for advanced information processing and storage.
\end{abstract}

\maketitle

Despite substantial advances in software neuromorphic computing based on mathematical models of synapses and neurons, a strong interest in the hardware implementation of neural networks remains. An effective hardware approach could offer faster and more energy-efficient solutions for complex tasks by using artificial intelligence algorithms. A key component in this paradigm is a device that emulates the plasticity of biological synapses, which are the interconnections between individual neurons. 

Many efforts to implement artificial synapses on chip are centered on memristors – passive two-terminal electrical devices, the conductance of which can be continuously modulated by external voltage pulses of appropriate amplitude/duration. Voltage pulses correspond to signals (“\textit{spikes}”) transmitted from a pre-neuron to a post-neuron, conductance mimics “synaptic weight”, that is transmission of spikes, and the conductance dynamics emulates synaptic plasticity. Memristors in which the conductance is adjusted entirely by external stimuli are known as “\textit{first-order memristors}”. With such memristors, by use of voltage pulses or certain voltage temporal profiles, a number of synaptic functions have been successfully demonstrated \cite{Nanoscale_memristor_device_as_synapse, An_electronic_synapse_device}. However, the first-order memristors does not possess internal temporal mechanisms determining its instant state and subsequent synaptic plasticity, which is the case in real biological systems. Indeed, the response of a biological synapse to a encoding spike train is determined not only by spike parameters, but also by the complex interplay of the multiple temporal processes in presynaptic and postsynaptic neurons, for example, $\mathrm{Ca^{2+}}$ ion diffusion across the postsynaptic membrane \cite{Ca_entry}. This effect provides the transient conditions for transmission adjustment on a short time scale, with or without long-term effects. Similar synaptic behavior can be achieved with a “\textit{second-order memristor}”, where the conductance is governed by both the external stimuli and its instant internal state. So far, the temporal dynamics of biological synapses has been achieved in filamentary \cite{Experimental_demonstration, Biorealistic_implementation}, bridge-based \cite{Memristors_with_diffusive_dynamics} and ferroelectric \cite{Mikheev} second-order memristors by appropriate internal decay processes, such as thermal dissipation of the oxygen vacancies in the conducting filament \cite{Experimental_demonstration}, the decay of the oxygen vacancies mobility \cite{Biorealistic_implementation}, the minimization of metal species interfacial energy \cite{Memristors_with_diffusive_dynamics} and dynamics of ferroelectric depolarization \cite{Mikheev}. The next step in the development of memristor-based neuromorphic systems involves the use of their cross-bar arrays, which has the inherent problem of unwanted leakage currents along the sneak paths, leading to the influence of memristor weights on each other. This is one of the major problems hindering the system implementation. To circumvent this problem, it has been proposed to use additional selector devices such as transistors and diodes \cite{Parallel_programming, Efficient} or to apply complex pulse inputs \cite{Fully_parallel}.

Alternatively, three-terminal transistor synaptic devices (so called “\textit{synaptic transistors}”) may be used to solve the sneak current path problem \cite{2D_floating_gate, Nonvolatile_Memory, Parallel_weight_update_protocol}. In such devices, a gate terminal controls and separates the weight updating and reading paths, canceling sneak currents. In addition, the absence of an additional selector device, which is required for a two-terminal-based synaptic array, helps to reduce the total chip area. To date, several synaptic transistors, including silicon \cite{Si_based_FET}, carbon nanotube \cite{Parallel_weight_update_protocol}, transition metal dichalcogenide \cite{Mimicking_neurotransmitter, MoS2_synaptic_device} and graphene \cite{Graphene_memristive_synapses} - based have been demonstrated. Among the above channels, graphene stands out for its transparency, flexibility, and high conductivity, which enables synaptic transistors for flexible chips, optoelectronic devices, and avoids the current reduction problem of device miniaturization, where scaling down makes the output signal too small for chip design \cite{Mikheev}. Chemical vapor deposition (CVD) technology of graphene growth enables the fabrication of large area single layers, making this material attractive to the industry. However, a graphene synaptic transistor based on all-technology-compatible materials has not been intensively studied in terms of its functionality as a synaptic device.

In this work, we demonstrate the highly adjustable conductance states and synaptic functions, such as pulse amplitude-dependent plasticity (PADP) and pulse duration-dependent plasticity (PDDP), paired-pulse facilitation/depression (PPF/PPD), spike rate dependent plasticity (SRDP), spike transition from short term plasticity (STP) to long term plasticity (LTP) in the graphene synaptic transistors that emulate biological synapses. A floating-gate transistor is designed with technologically compatible materials suitable for both solid-state and flexible electronics, and the effect of the floating-gate geometry on the memory window and thus on synaptic functionality is investigated.

\begin{figure*}[ht!]
    \includegraphics[width=1\linewidth]{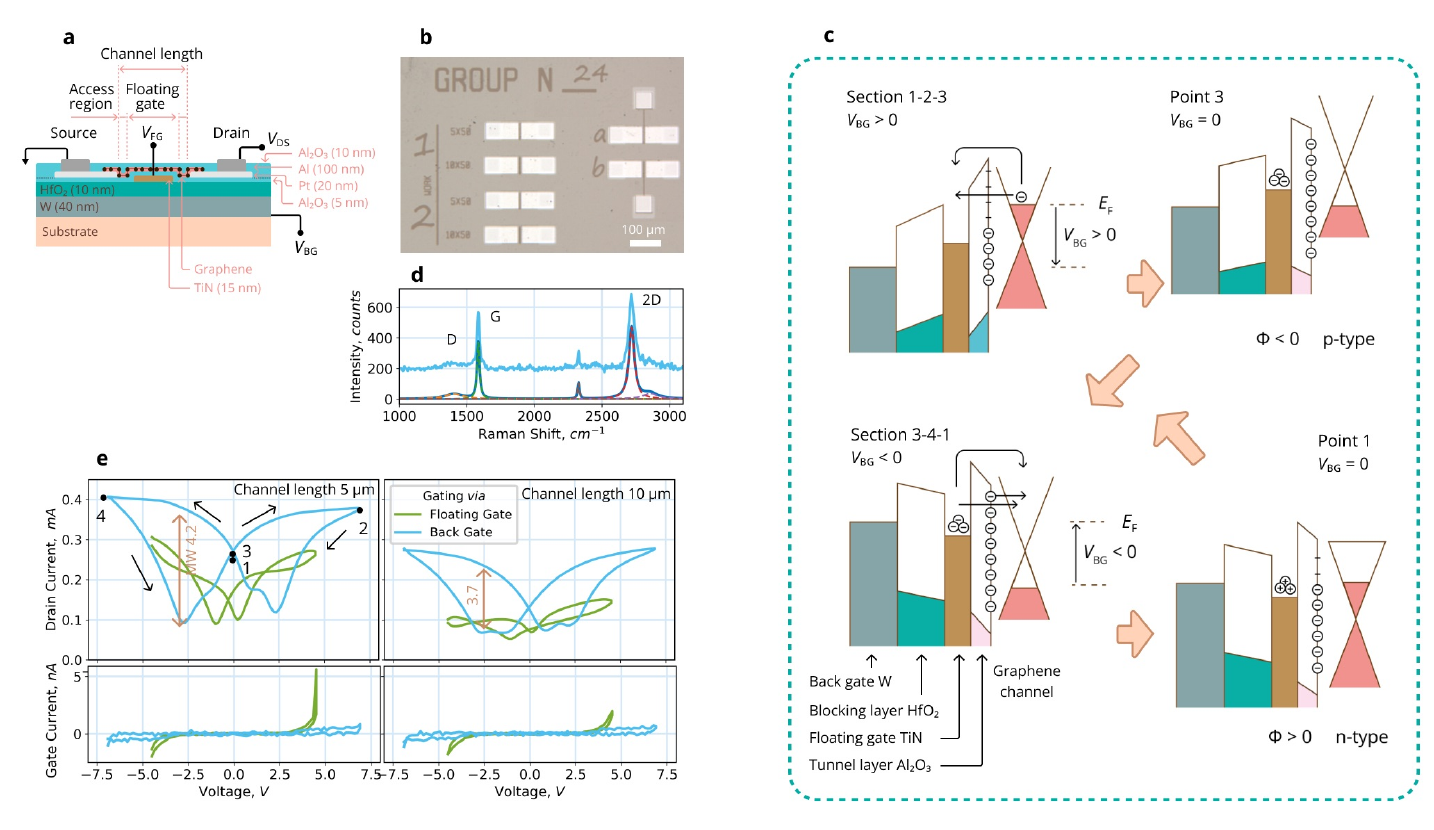}
    \caption{Working principle of synaptic transistor with CVD graphene channel. (a) Scheme of the transistor structure. (b) Optical microscope image of transistors with encapsulated floating gate and floating gate routed to the pads. (c) Energy band diagrams of the transistor during and after voltage application. (d) Raman spectrum of the CVD graphene. (e) Transfer characteristics and gate leakage curves of 5 and 10 $\mathrm{\mu m}$ channel length transistors that were encapsulated after removal of adsorbate from the graphene surface. Source-drain voltage equals to 50 mV.}
    \label{fig1}
\end{figure*}

The device is based on the layered structure metal (W)-isolator ($\mathrm{HfO_2}$)-metal (TiN)-isolator ($\mathrm{Al_{2}O_{3}}$)-graphene [\hyperref[fig1]{Fig.~\ref*{fig1}(a)}]. 40 nm thick W layer and 15 nm thick TiN layer grown by magnetron sputtering served as a back and floating gate, respectively. 10 nm thick $\mathrm{HfO_2}$ and 5 nm thick $\mathrm{Al_{2}O_{3}}$ grown by atomic layer deposition were blocking and tunnel layers. A 20 nm thick Pt film was deposited on top of the $\mathrm{HfO_2}$ layer using the e-beam evaporation and a $\mathrm{50 \times 100\,\mu m^2}$ drain and source were formed using optical lithography. To fabricate a graphene channel with a width of $\mathrm{50\,\mu m}$ and a length of 5-10 $\mathrm{\mu m^2}$, a monolayer graphene sheet, synthesized by CVD on copper foil, was transferred onto a substrate with pre-patterned electrodes using a wet transfer method. Excess graphene was removed by oxygen plasma etching following patterning by optical lithography. In order to ensure reliable electrical contact to the drain, source and floating gate, their contact pads were coated with a 100 nm thick aluminum film fabricated by e-beam evaporation. To protect the graphene channel from parasitic doping with adsorbates from the environment, the transistor structures were encapsulated with a 10 nm thick $\mathrm{Al_{2}O_{3}}$ film after preheating at $\mathrm{300^{\circ}C}$ for 3 h in Ar atmosphere required to remove the initial adsorbate layer. Finally, windows in the $\mathrm{Al_{2}O_{3}}$ film at the contact pads were made by etching using 5 \% HF buffer solution. Two types of floating gate transistors have been fabricated: those with access to the floating gate via contact pads and those with a fully encapsulated floating gate [\hyperref[fig1]{Fig.~\ref*{fig1}(b)}]. Electric characterization was made using the Cascade probe station coupled with the semiconductor parameter analyzer B1500A (Agilent Technologies) at standard conditions. Transfer, output and gate characteristics were measured using B1500A source measurement units, while synaptic functions were measured using waveform-generator fast measurement units.

\hyperref[fig1]{Fig.~\ref*{fig1}(d)} shows the Raman spectrum of the graphene channel, taken at a wavelength of 473 nm. The peak at approximately 1600 $\mathrm{{cm}^{-1}}$ is known as the G-band and corresponds to C–C bond stretching in $\mathrm{sp^2}$ carbon materials. The existence of a strong peak at a Raman shift value of 2500–2800 $\mathrm{cm^{-1}}$ indicates the presence of single-layer graphene, with the peak itself being referred to as the 2D-band. Notably, the Raman spectrum has a weak peak at approximately 1400 $\mathrm{{cm}^{-1}}$, known as the D-band. It corresponds to disorder in the $\mathrm{sp^2}$ structure of graphene and typically manifests itself in CVD-grown graphene.

The floating gate was designed so that its potential affects the conductivity of only part of the channel. The floating gate width was equal to the graphene channel width (50 $\mathrm{\mu m}$) and the gate length was 3 $\mathrm{\mu m}$, while the channel lengths were 5 and 10 $\mathrm{\mu m}$ for the two types of transistors. Thus, the area between the source/drain and the floating gate (access region) was 1 and 3.5 $\mathrm{\mu m}$ for transistors with channel lengths of 5 and 10 $\mathrm{\mu m}$, respectively. This device design leads to a split Dirac point in the transfer characteristics of the transistor (blue lines at the top of \hyperref[fig1]{Fig.~\ref*{fig1}(e)}). The splitting is more prominent in transistors with 10 $\mathrm{\mu m}$ channel length, where the length of the floating gate is almost equal to the length of the access region. By design, the transfer characteristics have hysteresis, which is usually attributed to the influence of the charge trapped by the adjacent oxide layer \cite{Hysteresis}.

The physical mechanism of Dirac point splitting is related to the presence of two types of channel regions: above the floating gate and above the back gate (access region). Indeed, the device has two reservoirs for the charge flowing when a voltage is applied to the back gate: traps at the interface between the $\mathrm{Al_{2}O_{3}}$ film and graphene and the floating gate itself. During the application of a positive voltage, electron injection occurs due to tunneling through the ultrathin $\mathrm{Al_{2}O_{3}}$ layer or thermionic emission  [\hyperref[fig1]{Fig.~\ref*{fig1}(c)}]. The injected electrons are captured by the traps and the floating gate. Due to the excess negative charge, these reservoirs acquire a negative potential, which causes p-doping of graphene and a shift of the Dirac point to the positive voltage region the larger the reservoir potential. The splitting of the Dirac point is due to the different potentials of traps and the floating gate. During the application of a negative voltage, both excess and intrinsic free electrons are removed, resulting in positively charged reservoirs. The Dirac point shifts to the region of negative voltages, which is also accompanied by its splitting. It is worth noting the symmetry of the transfer characteristics, indicating the successful removal of adsorbate prior to encapsulation of the transistor and, hence, the absence of the built-in electric field.

When the voltage is applied directly to the floating gate, Dirac point is naturally not splitted (green lines at the top of \hyperref[fig1]{Fig.~\ref*{fig1}(e)}). In this case, the floating gate current $\mathrm{\textit{I}_{FG}-\textit{V}_{FG}}$ curves show a significant injection current through the tunneling layer to the gate (bottom plots in \hyperref[fig1]{Fig.~\ref*{fig1}(e)}), and the transfer $\mathrm{\textit{I}_{DS}-\textit{V}_{FG}}$ characteristics are significantly distorted (green lines at the top of \hyperref[fig1]{Fig.~\ref*{fig1}(e)}). 

The Dirac point splitting causes the increase of the memory window, which becomes close to its limit for graphene at room conditions. The memory window (MW) reaches 4.2 and 3.7 for transistors with channel lengths of 5 µm and 10 µm, respectively [\hyperref[fig1]{Fig.~\ref*{fig1}(e)}]. Since the performance of encapsulated transistors is highly stable at room conditions and the drain current magnitude is perfect for chip design, this memory window is sufficient for application in electronics.

\begin{figure}[ht!]
    \includegraphics[width=1\linewidth]{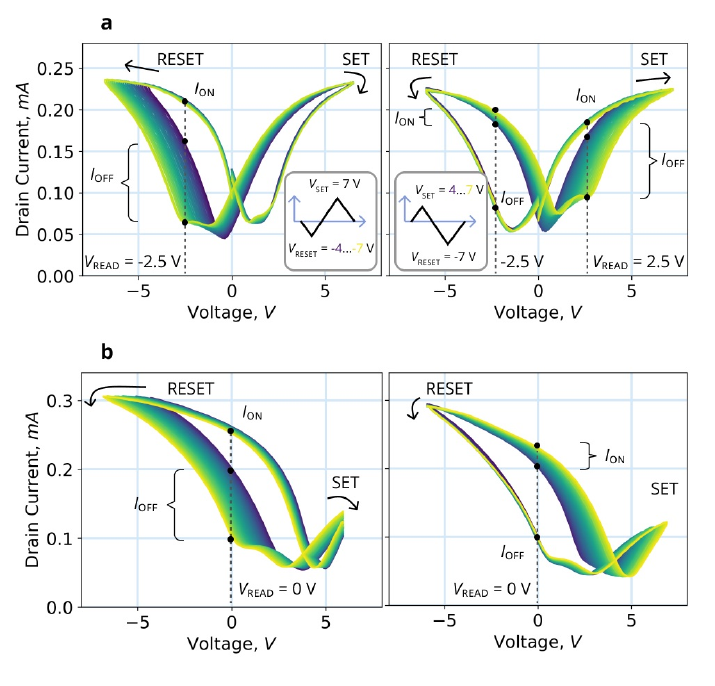}
    \caption{Memristive properties of transistors fabricated (a) with and (b) without adsorbate removal. Source-drain voltage equals to 50 mV.} 
    \label{fig2}
\end{figure}

As mentioned above, transfer characteristics are well symmetrical, indicating the successful removal of adsorbate prior to encapsulation of the transistor. However, for practical applications, symmetrical transfer characteristic is not optimal, because it implies reading information when the gate voltage is applied, e.g. $-3$ V for the structure with channel lengths of 5 $\mathrm{\mu m}$ [\hyperref[fig1]{Fig.~\ref*{fig1}(d)}]. If the synaptic transistor is designed so that the maximum memory window is at zero gate voltage, this would reduce the power consumption of the device. This can be achieved by using the so-called doping of graphene with adsorbate from the atmosphere. This phenomenon consists in a shift of the Dirac point to the positive voltage region in graphene stored under standard conditions. The shift is related to the electrostatic contribution of the adsorbate charge. In order to exploit it, synaptic transistors were fabricated in such a way that graphene was encapsulated with aluminum oxide without the procedure of long preheating of the sample in argon. From the comparison of \hyperref[fig2]{Fig.~\ref*{fig2}(a)} and \hyperref[fig2]{Fig.~\ref*{fig2}(b)}, it can be seen that this approach does allow to control the position of the Dirac point.

The memristive behavior of both transistors with symmetrical and shifted transfer $\mathrm{\textit{I}_{DS}-\textit{V}_{BG}}$ characteristics is confirmed by a number of resistance states observed on the characteristics [\hyperref[fig2]{Fig.~\ref*{fig2}(a, b)}]. Virtually continuous range of resistance levels between low and high resistance states were obtained by changing the RESET voltage amplitude between $-4$ and $-7$ V or the SET voltage amplitude between 4 and 7 V. The conductance tuning occurs due to a gradual change in the potential of the floating gate and interface traps due to the gradual increase in charge injected and pulled by the applied voltage [\hyperref[fig1]{Fig.~\ref*{fig1}(c)}].

As can be seen in \hyperref[fig2]{Fig.~\ref*{fig2}(a, b)}, for a transistor with a symmetric transfer characteristic, the gate read voltage should be about $\mathrm{\textit{V}_{READ} = -2.5\,V}$, where the memory window is maximized. For the transistor with shifted transfer characteristic, the gate read voltage is $\mathrm{\textit{V}_{READ} = 0}$. At these read voltages, the RESET procedure results in a decrease in the readout drain current, while the SET procedure results in its increase. It is noteworthy that for a transistor with a symmetrical characteristic it is possible to use a gate read voltage of the opposite polarity $\mathrm{\textit{V}_{READ}  = 2.5\,V}$ (right in \hyperref[fig2]{Fig.~\ref*{fig2}(a)}). Then the SET and RESET voltages also change polarity.

\begin{figure*}[ht!]
    \includegraphics[width=1\linewidth]{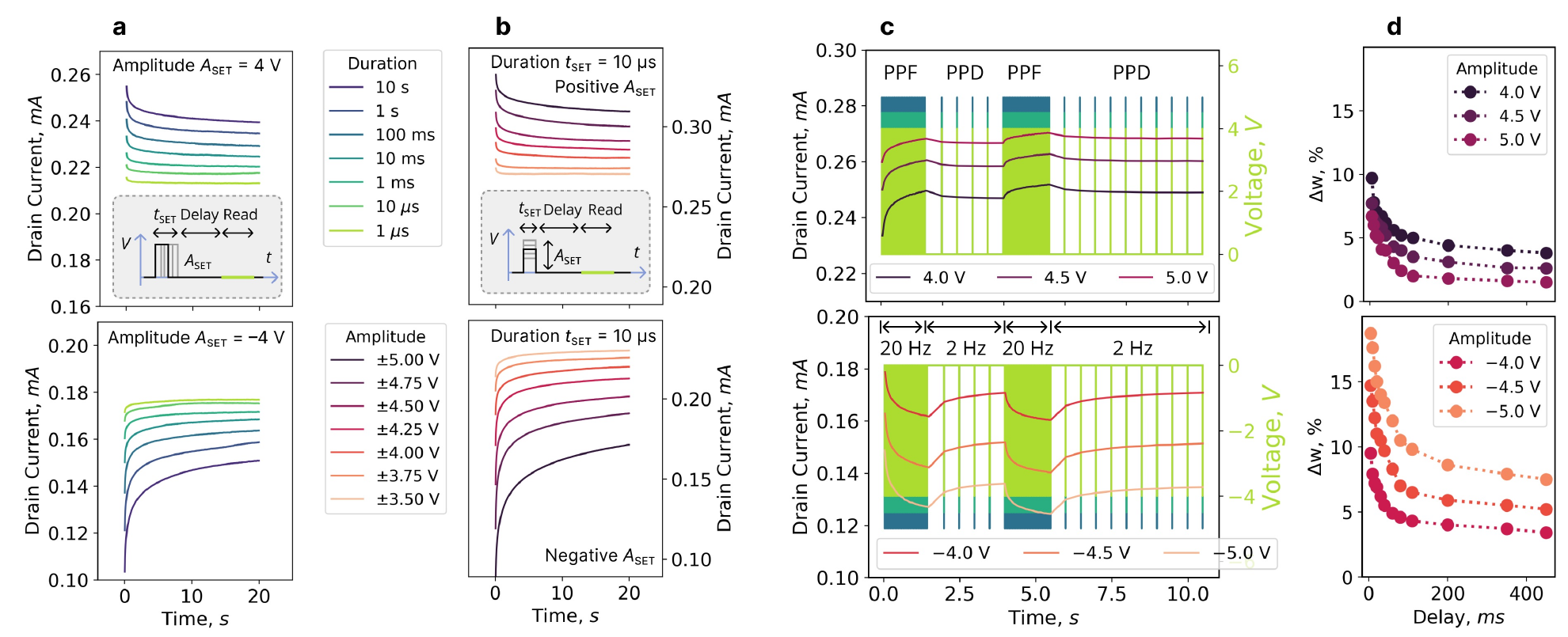}
    \caption{ Synaptic functionality. (a) Pulse amplitude-dependent and (b) pulse duration-dependent plasticity. (c) Spike rate dependent plasticity and associated paired-pulse facilitation/depression. (d) The conductance (synaptic weight) $\mathrm{\Delta w}$ change as a function of delay time between pulses.}
    \label{fig3}
\end{figure*}

To investigate the synaptic functionality of the artificial transistor synapse, we focused on a transistor with asymmetric transfer characteristic, which is more promising due to lower power consumption. First of all, the temporal dynamics of the synapse weight was investigated after a switching voltage pulse was applied to it. For this purpose, we measured the temporal current response after single pulses of fixed amplitude (of both polarities), but varying durations and after single pulses of fixed duration, but different amplitudes (of both polarities). The drain current was measured at a gate voltage $\mathrm{\textit{V}_{READ}  = 0}$ and source-drain voltage $\mathrm{\textit{V}_{SD}  = 50\,mV}$. \hyperref[fig3]{Fig.~\ref*{fig3}(a, b)} show that after switching, the drain current gradually relaxes to some stationary value corresponding to long-term memory. The relaxation process itself is a manifestation of short-term memory and reflects the pulse amplitude-dependent plasticity (PADP) and pulse duration-dependent plasticity (PDDP) synaptic functions, indicating the biorealistic behavior of the fabricated synaptic transistors. Such temporal dynamics is a signature of a second-order memristor, but in advantageous three-terminal architecture. In this device, the intrinsic mechanism of conductance change is the charge state dynamics of the floating gate and interface traps. It is worth noting the asymmetrical current response to positive and negative pulses. Namely, negative voltages cause a wider range of initial conductivity values and more pronounced relaxation. This result is consistent with the $\mathrm{\textit{I}_{OFF}}$ and $\mathrm{\textit{I}_{ON}}$ ranges observed in the transfer characteristics and is related to their asymmetry. 

The fact that conductance relaxes to some stationary levels determined by the duration, amplitude, and polarity of the pulses reflects the transition from short term plasticity (STP) to long term plasticity (LTP). Long term memory corresponds to the capture of charge by deep levels at the graphene interface, whose effect is similar to the principle of flash memory. 

The PADP and PDDP synaptic functions indicate that the structure should be expected to have a frequency response of conductance to the spike train, thus emulating frequency response of biological synapses. Indeed, the device demonstrates spike rate dependent plasticity (SRDP) and, similar in its physical origin, paired-pulse facilitation/depression (PPF/PPD) [\hyperref[fig3]{Fig.~\ref*{fig3}(c, d)}]. In biology, at low frequency of the spike train, the synaptic weight does not undergo long-term changes, whereas the higher frequency leads to the increase in the synaptic weight. Therefore, the specific feature of the biological synapse is its own response to the frequency of the spike train encoding the information. In a solid-state second-order memristor and synaptic transistors, the conductance is similar to the synaptic weight. The positive biasing at high (20 Hz) frequency gradually increase a negative potential of the floating gate and interface traps setting the structure in the lower resistance state (ON). This functionality is similar to the paired-pulse facilitation (PPF) in biological synapses. The decrease in the pulse train frequency gradually turns the device in the high resistance state (OFF) due to the gradual emission of the trapped charge back to the graphene, thus emulating the paired-pulse depression (PPD) [\hyperref[fig3]{Fig.~\ref*{fig3}(c)}]. It should be noted that the frequency response of the memristor is effectively tuned by varying the amplitude of switching pulse [\hyperref[fig3]{Fig.~\ref*{fig3}(a)}], which enables additional possibilities in hardware neuromorphic computing. The frequency response is also different for positive and negative pulses. Conductance dynamics is more pronounced for negative voltage, which is due also to the asymmetry of the transient characteristics.

Meanwhile, two other more fundamental differences in the frequency response to positive and negative spikes are worth noting. First, although high-frequency excitation by positive spikes causes facilitation of signal transmission, high-frequency negative spikes cause depressed signal transmission, i.e., the opposite effect. To accurately emulate the behavior of a biosynapse, one should use unipolar transistor operation, for example, one could use only positive gate voltages for conductance switching. However, the second mode of operation expands the possibilities for hardware implementation of neuromorphic computation, making it more flexible. The second notable result is that as the amplitude of positive high-frequency spikes increases, the relative change in conductance slows down, reaching saturation. As the amplitude of negative spikes increases, facilitation only increases. This effect is probably due to the asymmetry of the potential barrier and current transport from and to the floating gate.

In conclusion, a technology-compatible synaptic transistor based on a graphene channel and a high-k gate stack was designed, engineered and fabricated. It exhibits a virtually continuous range of multiple conductance levels. Engineering of the device geometry and graphene surface state allowed maximizing the memory window and minimizing power consumption. The synaptic transistor has synaptic plasticity and exhibits a number of synaptic functions, such as paired-pulse potentiation/depression, spike-rate-dependent plasticity and others. The implemented synaptic transistor can serve as a building block for the development of neuromorphic computing architectures implemented on chip.

\subsection*{Acknowledgements}
The work was performed using the equipment of MIPT Shared Facilities Center. Fabrication, designing and synaptic characterization were supported by the Russian Science Foundation (Project No. 20-19-00370, \href{https://rscf.ru/en/project/20-19-00370/}{https://rscf.ru/en/project/20-19-00370/}). The development of methodology and software for transistor characterization measurements were supported by the Ministry of Science and Higher Education of the Russian Federation (agreement no. 075-03-2024-117, project No. FSMG-2022-0031).

\subsection*{Data Availability Statement}
Data available in article.

\subsection*{References}
\bibliography{arXiv}

\end{document}